\documentclass{article}
\usepackage{amsmath}
\usepackage{amscd}
\usepackage{amsthm}
\usepackage{amssymb} \usepackage{latexsym}
\usepackage{eufrak}
\usepackage{euscript}
\usepackage{epsfig}
\usepackage{graphics}
\usepackage{array}
\usepackage{enumerate}

\usepackage{amssymb}

\newcommand{\be}{\begin{equation}}
\newcommand{\ee}{\end{equation}}
\newcommand{\ba}{\begin{eqnarray}}
\newcommand{\ea}{\end{eqnarray}}
\newcommand{\rf}[1]{(\ref{#1})}
\newcommand{\bi}{\bibitem}

\newcommand{\bel}[1]{\begin{equation}\label{#1}}
\newcommand{\qe}{\end{equation}}

\title{Learning, evolution and population dynamics}
\author{J\"urgen Jost\footnote{This work was partially supported by a grant from the Volkswagenstiftung.} and Wei Li\\
Max Planck Institute for Mathematics in the Sciences,\\
Inselstr.22, 04103 Leipzig, Germany}

\begin{document}

\maketitle

\begin{abstract}
We study a complementarity game as a systematic tool for the investigation of the interplay between individual optimization and population effects and for the comparison of different strategy and learning schemes. The game randomly pairs players from opposite populations. The game is symmetric at the individual level, but has many equilibria that are more or less favorable to the members of the two populations. Which of these equilibria then is attained is decided by the dynamics at the population level. Players play repeatedly, but in each round with a new opponent. They can learn from their previous encounters and translate this into their actions in the present round on the basis of strategic schemes. The schemes can be quite simple, or very elaborate. We can then break the symmetry in the game and give the members of the two populations access to different strategy spaces. Typically, simpler strategy types have an advantage because they tend to go more quickly towards a favorable equilibrium which, once reached, the other population is forced  to accept. Also, populations with bolder individuals that may not fare so well at the level of individual performance may obtain an advantage towards ones with more timid players. By checking the effects of parameters such
as the generation length or the mutation rate, we are able to
compare the relative contributions of individual learning and
evolutionary adaptations.

{\bf Keywords: Evolutionary Complementarity Game; Individual
Learning; Population Dynamics; Evolutionary Adaptation}

\end{abstract}

\section{Introduction}
Our first aim is to investigate the relation between individual optimization and
the resulting collective dynamics in an evolving environment. This topic has a long
history, starting (at least) with Mandeville's essay \cite{Man} on bee colonies and
Adam Smith' \cite{Smi} invisible hand. From a scientific point of view, it is important
to analyze the validity and generality of the many claims that have been brought forward
since then and to identify their necessary and sufficient assumptions. For that purpose,
we need a simplified formal model in which one can isolate the key mechanisms and features
without all the contingent details of real world situations. In order to proceed in that
direction, we utilize an agent based model that can be readily simulated and is also amenable
to formal analysis. This model is a population game where the equilibrium at the individual
level is degenerate so that the selection among the possible equilibria results from the
collective dynamics at the population level. \\
Our second aim is to compare the strength of different learning
schemes in an evolving competitive situation, that is, where the
opponents also try to learn  efficiently. For this purpose, in the
sense of statistical learning theory, every agent needs to have a
stream of stochastic input data on the basis of which he can develop
his models. In order to overcome the limitations of classical game
theory, an agent encounters in each round an opponent that is
randomly chosen from an opponent population. When all agents in one
population employ a particular learning strategy, and all agents
from the opponent population
employ another strategy, we can then see in favor of which population the  equilibrium develops. \\
Our third aim is to connect and compare the two previous aspects, evolution and learning.
Thus, we want to see what is better for agents and for populations of agents, to adapt by
evolution or to learn by individual experience. \\
Our model is the following complementary game played between members of opposite populations,
as introduced in \cite{JostLiFirst}: A buyer and a seller meet and independently each make an
offer between 0 and $K$ ($K$ here is a sufficiently large integer, usually taken to be 50 in our simulations).
When the buyer's offer $k_b$ is at least as large as the seller's offer $k_s$, a deal is concluded
and the buyer gains $K-k_b$, the seller $k_s$; otherwise, they gain nothing. Thus, in order to be
most successful, the buyer should offer not less than the seller is asking, because otherwise
he will not get a deal, but also not much more, in order not to pay too much. We note that every
integer between 0 and $K$ is an equilibrium in the sense that no player can do any better than
playing that value if his opponent does so. This game then is played repeatedly between members
of two opposing populations of the same size, the buyers and the sellers. In each round, the
members of the two populations are randomly paired, that is, every buyer is paired with a randomly
chosen seller. After a fixed number of rounds, the accumulated gains of the agents in each population
are compared, and on the basis of this fitness function, some evolutionary scheme constructs a new population. \\
Thus, the basic situation is symmetric between the two players, the buyer and the seller, and
also between the two populations. We can then break that symmetry by equipping the buyers and the
sellers with different strategy spaces. The differences here could simply be differences of memory
span, that is, how many encounters an agent can remember and utilize for determining his own
current bid. The agents could also employ totally different strategies. Here, the possible strategies
could range from playing a random number to an elaborate scheme for computing a bid on the basis of all
information available from the agent's experience or even including the experiences of his friends in
his own population. \\
In general, the agents of each population will adapt through individual learning and through fitness
based evolution. Thus, even if the agents in a population act completely independently of each other,
they will feel the long term effects of the actions of their fellows through the collective adaptation
of the agents of the other populations. \\
If the members of one population, say the sellers, could coordinate their actions, their best strategy
would consist in always choosing that bid that is optimal for them, $K$ in this case. The other population
would then have no choice but to accept that and also play the same bid. However, as long as the buyer
population has not evolved to that state that is so unfavorable to them, it pays for any individual
seller to lower his bid and to increase his chance for a successful deal. Thus, such an agent would
be more successful than the ones keeping to the population optimum $K$, and because of his higher fitness,
his strategy would then be more frequently represented in the next generation. Thus, in this evolving scheme,
the group optimum is not stable against defections in the own population. In particular, those agents in a
population that are individually fittest can cause a decrease of the fitness of the population as a whole.
This relates to our first aim.\\
Also, since the members of both populations are trying to maximize their fitness, one cannot expect
that either population can enforce that equilibrium that is optimal for itself onto the other population.
In a symmetric situation, we would expect $K/2$ as the eventual steady state. When the buyers and the sellers
employ different strategies, we can simply decide which strategy is superior by checking whether the steady
state reached in that situation is smaller or larger than $K/2$. In the first case, the buyers are doing better,
in the second, the sellers. This then allows us to address our second aim.\\
We can also play with such parameters as the generation length or
the mutation rate in the evolutionary step. In that way, we can
compare the relative contributions of individual learning and
evolutionary adaptations,
as formulated in our third aim. \\
Although the rules of our game are extremely simple, the action takes place at three different levels:
The individual agents evaluate the information they obtain from their interactions and use that to
compute their next own actions (first level, information evaluation and learning), they compete with
each other inside a population (second level, adaptation and evolution), and the populations are compared
with each other (third level, competition between strategy spaces). The link between the first two levels
is provided by the fitness function, the one between the higher levels by the collective dynamics at the
population level resulting from the individual optimizations.\\\\
While what we present here, clearly does not yet constitute a complete theory, we believe that we have found
a formal model that on one hand is simple enough for easy simulations and also permits formal analysis, but
on the other hand is rich enough to capture many of the essential features of the conceptual problems we wish to address. \\
We shall first put our model into the perspectives of game
theory and of certain traditions in economics. After that, we shall start with some mathematical considerations before we present various simulations that both illustrate
some of our formal reasoning and yield insightful results beyond those that we can demonstrate formally.\\
We thank a referee for  constructive comments.

\section{Game theory}
In order to put the present work into perspective, we now shall
discuss in more detail how it fits into modern game theory. In the
classical model of game theory as introduced by von Neumann and
Morgenstern\cite{vNM}, we have two players that meets once. They have a finite
set of action options, and the pay-off for each of them is determined
by her own and her opponent's action.  They are both perfectly rational
and possess and can utilize all relevant information. Thus, they both try to maximize their
pay-off function, each also knowing her opponent's pay-off function
and therefore choosing that action that best anticipates the
opponent's move that is assumed to be in the same way anticipating. In
such a situation, there exists a Nash equilibrium\cite{Nash} in which no player
can improve her pay-off by changing her action, given that the
opponent will react correspondingly in her own best interest. That
Nash equilibrium need not be unique; in fact, in our basic model game,
each value $k$ between 0 and $K$, when played by both players, is a
pure strategy Nash equilibrium. In addition, there are mixed strategy
equilibria.  In particular, there is no rational way for a player
to decide which bid to play because each bid is a pure Nash equilibrium
when also played by her opponent. \\
We are therefore interested in the mechanisms that can select between
all those equilibria. Since in our model, the game is played
repeatedly, and the agents can benefit from their own experience (or
those of other players in some version of our game), this brings us to
the theory of learning in games, see e.g. \cite{FuLe} as a reference
for our discussion of this topic. Also, the game is played in
populations of agents, which leads us into evolutionary game theory
where we can use, e.g., \cite{Wei} as a reference. Leaving the issue
aside for the moment that in our model we have two distinct
populations, we are considering a random-matching model where in each
round, all players are randomly matched, but can observe only their
own matches. Also, in our model, no player acts in the interest of the
population, but only myopically pursues her own aims. Concerning
learning and evolution, we have a case of fictitious play where
players only observe (and perhaps memorize) the results of their own
matches. (Nevertheless, we shall also consider scenarios where players
have information about the performance of selected other players in
their own population, their ``friends''.)  The players then evolve
after many rounds of play  according to their relative fitness in
their population (as in evolutionary algorithms), in contrast for
example to replicator dynamics where the relative frequencies of
strategies continuously change according to their actual performance
(see \cite{HoSi}). In fact, we shall discuss below the issue of
generation length, i.e., after how many rounds the fitness of the
players is evaluated; the extreme case of generation length 1 could be
made to correspond to discrete time replicator type dynamics. Such an
on-line evolutionary adaptation, however, would prevent individual
agents from improving their performance on the basis of their own
experience, that is, learning. In other words, we are interested in a
hybrid of games with learning and evolutionary games, in order to
assess the relative strengths and problems of the two schemes.  \\
Coming to evolutionary aspects, the concept of an evolutionarily
stable strategy (ESS) as introduced by Price and Maynard Smith \cite{MaPr}
is not directly applicable to our setting because the matches are played between
members of different populations. (Also, as emphasized for instance in \cite{VR}, the concept of an ESS takes as its base situation a monomorphic population, that is, one where all members utilize the same strategy. In many applications, however, one is naturally interested in the stability of a polymorphic population against invasions of mutants. In our case, however, the basic equilibria that we shall take as our default situation do consist of monomorphic populations applying pure strategies.)  Versions of evolutionary stability for multipopulation games have been developed in \cite{Tay,Cr}, for instance (see \cite{Wei} for further references). In those definitions, a population with a rare  mutant in one of the populations is compared with the original population in its performance against the other unchanged populations.  By the result of \cite{Se}, a strategy is evolutionarily stable iff it is a strict Nash equilibrium.  In particular, when the game is at a
pure Nash equilibrium, any mutant strategy performs less well than the
dominant strategy -- {\it provided the opposite population does not
  adapt to those mutants.}  This provision shows that the concept of evolutionary stability is essentially a static one and therefore not so well suited for our setting. Whether it will be advantageous for members
of a population to adapt to a rare mutant in the opposite population, however, depends on the relations between
the various parameters as the following heuristic reasoning
shows. When the generation length is short, that is, each player plays
only few rounds before her fitness is evaluated, then it will not pay
off to adapt to a possible rare mutant in the opposite camp. Namely, in
that situation, a player is unlikely to encounter such a rare mutant
in her lifetime, and there is therefore no point in playing a strategy
that is most likely to be inferior in all encounters. When the number
of matches played becomes higher, comparable to the size of the
population, then there is a substantial chance to encounter a rare
mutant at least once. Let us assume that the player is a buyer, and
the rare mutant is a seller that asks a higher price, say $k+1$, than the
population equilibrium $k$. When the buyer offers $k+1$ in all $m$
matches, her accumulated gain is $m(K-k-1)$. When she offers only $k$ and
encounters the mutant once, she will accumulate $(m-1)(K-k)$. So,
adapting to the mutant will be advantageous when $m<K-k$. As a typical
evolutionary outcome of our game is that $k \sim K/2$, this means that
evolutionary stability will depend on the population size. In the
preceding heuristics, we have assumed that there is only a single
mutant. If there are more of them, of course, they can destabilize the
reaction of the opposite population more easily. In any case, we arrive at the heuristic conclusion that in a sufficiently large population, invasions of rare mutants should not lead to a response adaptation of the other population that ultimately makes the performance of the mutants superior. This is, of course, in accordance with the general idea of an evolutionarily stable strategy. --  The issue of stability of strategies in multipopulation games can, however, also be addressed by the inherently dynamical framework of replicator dynamics. This approach has been developed in \cite{AmHo,HoSi1,Wei}. In particular, in our game, mixed Nash equilibria are no longer (Lyapunov or asymptotically) stable for the replicator dynamics, as follows from the analysis of \cite{JostLiFirst}. The pure Nash equilibria, however, remain stable attractors, and it depends on the initial conditions which one is ultimately approached. This fits, of course, with the general theory. A strategy combination of the two populations is evolutionarily stable iff it is a strict Nash equilibrium (which necessarily is pure) (by \cite{Se}, as already mentioned) iff it is asymptotically stable for the standard replicator dynamics, see \cite{HoScSi,HoSi1,Wei}. In a similar direction, we have the regular and pay-off monotonic evolutionary systems of \cite{VR}, a replicator type dynamical model of multi-population games, whose asymptotically stable states also yield Nash equilibria.\\
In summary, in the context of game theory, the distinctive features of our model are the following. A-priori rational reasoning as in the standard game theoretic paradigm does not lead to a unique solution because there exist multiple equilibria. Which one is achieved depends on the historical contingencies of the evolution process. Those in turn are governed by the available strategic options for utilizing the information obtained by repeated interactions with different opponents. Since we have a 2-population game, we can equip the two populations with different strategy spaces, in particular different learning schemes, and can then see which one fares better. In other words, the game theoretic degeneracy of our model allows us the comparison of evolutionary and learning mechanisms that can break that degeneracy favorably.  In our game, players can have full information about the pay-offs of themselves and their opponents, but this information is not so helpful as the population dynamics are not only shaped  by their own actions and the responses of their opponents, but also by the ones of their conspecifics with whom the compete via their fitness function. The players, however, do not have information about the previous actions of their current opponents, but only information about the actions of their previous opponents in their past encounters with them. We shall also investigate the situation where a player also has information about the behavior of the opponent's of some other members of her own population, her ``friends'', but since in any case the opponents are randomly sampled, this will not lead to decisive advantages in our evolutionary simulations. Thus, for instance, some of the learning schemes analyzed in \cite{VR} are not applicable. In particular, a player cannot distinguish between her different opponents and so there is no way how a player could respond to her opponent individually.

\section{Economic ideas about the evolution of institutions}
While we have mentioned Adam Smith in the introduction, a conceptually closer starting point for our model is the work of Carl Menger\cite{Me}, the founder of the Austrian school of economics, with his (direct or indirect) disciples F.von Wieser, E.von B\"ohm-Bawerk, L.von Mises, F.von Hayek and J.Schumpeter (see e.g. \cite{Schum}). For Menger, economic phenomena were the emergent product of individual rational actions in situations where information is meagre and where actions are costly. Methodological individualism kept Menger and his school critical of techniques of aggregations as they underly neoclassical economics (see e.g. the standard reference \cite{Sa}) and of ideas of central planning or welfare economics, and emphasis on rational actions clearly distinguished them from philosophies like utilitarism. More recent work on the evolution of institutions as the individually unintended collective result of individual actions that are rationally optimizing their own target functions in the presence of limited information includes  \cite{Sch,Sug,Nor}. For a pertinent case study, see for instance \cite{Gre}. This approach has been cast into the framework of game theory by P.Young\cite{You}. Young's setting makes the assumptions that players are randomly drawn from large populations, that their probability of interactions depends on exogeneous factors like spatial proximity, and that they try to act rationally on the basis of the limited information that they have available. Whereas the second point is not elaborated upon in our model, the other two ones constitute also a basis for our contribution. We should point that in our model, there is no room for bargaining. While this restricts its applications to certain economic situations, it is essential for simplifying the strategy space available to the individual players so that their rational strategy choice behavior can be investigated more easily. Nevertheless, our scenario can be extended to include more complex forms of social interaction, like punishment \cite{Li}.\\
Also, Young\cite{You} considers the effects of random perturbations caused by unpredictable exogenous events or agent behaviors. Although we do not elaborate upon all these aspects here, this can also be naturally incorporated into our model, and of course, it also relates to the discussion of the stability concepts in the formal analysis, see \cite{FoYou,You1}.

\section{Adapting to random distributions}
Let $k_b(n)$ and $k_s(n)$ be the bids of a buyer and a seller that
interact at time $n \in {\mathbb N}$. When the seller chooses his bid $k$ with
probability $p(k)$, then his expected gain at time $n$ is
\bel{1}
\sum_{k=0}^K k\ p(k)\ p(k \le k_b(n))
\qe
because he gains $k$ when his bid is not larger than the one of the
buyer he encounters. \\
When the populations are sufficiently large (compared to the
generation length), the chance that our seller encounters the same
buyer repeatedly is small and can be neglected. In particular, this
implies that the buyer's bids will not reflect previous bids of that
seller, but rather constitute a response to the action of other
sellers that are randomly sampled from the population. Assuming that
our seller is not in contact with his fellow seller, the action of the
buyer will then be random for him, that is, $k_b(n)$ can be treated as
a random number. The underlying distribution will in general not be
stationary in time because the buyers may also learn from their
previous encounters and adapt their bids correspondingly.\\
\rf{1} leads to the following simple observation. When we consider
$k_b(n)$ as a random variable, the seller should find that response
$0\le k_s \le K$
that satisfies
\bel{2}
k_s=\text{argmax } k\ p(k \le k_b(n)).
\qe
When that value is unique, it is the best strategy to always play that
number because \rf{1} is a convex combination of the different $k$
values since $\sum_k p(k)=1$ as we are dealing with
probabilities. When that value is not unique, he can choose any convex
combination of the maximizers. \\
Given the probability distribution for $k_b(n)$, we can then determine
the optimal bid $k_s$ (of course, the crucial point is that this
probability distribution for the buyer bids is not known to the
seller, and so, in general, he will not be able to identify that
optimal value, but nevertheless, it is instructive to compare possible
strategies with the optimal one). When the distribution for $k_b(n)$
is a Dirac distribution with the value $k_0$, that is, when the buyer
always selects the same value $k_0$, then obviously the seller's best
choice is to utilize the same value. When the values of $k_b(n)$ are
distributed uniformly, that is, $p(k_b(n))=\frac{1}{K+1}$ for
$k_b(n)=0,\dots ,K$, then the expected gain for the seller's choice $k$ is
$\frac{k(K+1-k)}{K+1}$ whence the optimal value is $k=\frac{K+1}{2}$
which leads to the expected gain $\frac{K+1}{4}$. This then is better
than any other response. For example, if the seller also chooses his
bid $k$ randomly with the same uniform distribution, his expected gain
is
\bel{3}
\frac{1}{K+1}\sum_{k=0}^K k \ p(k \le k_b(n))=\frac{1}{K+1}\sum k \frac{K+1-k}{K+1}=\frac{K(K+2)}{6(K+1)}
\qe
which is smaller than $\frac{K+1}{4}$. This also applies when the
seller chooses as his bid at time $n$ his opponent's bid $k_b(n-1)$
from the previous round. The reason is that we are assuming here that
the buyer's bid is random, and so, when following the buyer's previous
bid, the seller simply takes a random value for $k$. When, in
contrast, the seller averages the buyers' previous bids, that is,
chooses
\bel{4}
k_s(n)=\frac{1}{n-1}\sum_{i =1}^{n-1}k_b(i)
\qe
(where $k_b(i)$ is the value of his opponent encountered at time
$i$), then, when $n$ gets large, by the central limit theorem,
$k_s(n)$ approaches the mean value of $k_b(n)$, assuming that the
underlying distribution is stationary. When that distribution is
uniform, we end up with the optimal value $k=\frac{K+1}{2}$. So, at
least in this simple case, we can deduce that averaging over previous
bids is a better strategy than simply taking the value encountered in
the last step. (This argument generalizes to certain (but not all)
more general distributions for the buyer bids.)\\
We also see the following. When both seller and buyer start with the
same random distribution for their bids and adopt the strategy to copy
the previously encountered opponent's bid, then they will stay with
that same random distribution forever. Thus, no progress is made in
that case. This is somewhat analogous to persistent miscoordination in iterated two-player games, see e.g. \cite{FuKr}. In fact, in our game, copying the previous opponent's bid is the rule that follows from Cournot adjustment, that is, from always choosing the best response to the previous round, see \cite{FuLe}. --  When, in contrast,  both players average, then they are both
expected to end up with the same value $k=\frac{K+1}{2}$ when the time
$n$ is sufficiently long. Also, when they both try to optimize
according to \rf{2}, even though they do not know the other's strategy,
they are expected to end up both with the same fixed value which when
they both act optimally will be the value $k=\frac{K+1}{2}$ which we
had already found to be the optimal against a uniform random
strategy. But this already leads us to the issue of

\section{Learning}
In our setting here, learning consists in using the experience from
previous encounters to determine the bid for the present round. The
problem then is to utilize the information from those previous
encounters in the most efficient manner, without wasting too much
effort with useless or disadvantageous trials. Since both
populations are adapting, we have a more subtle situation than simply
trying to learn an unknown, but fixed probability distribution as for
example in statistical learning theory \cite{Vap}.\\
In principle, the
learning strategy should compute the actual bid as a function of all
previously encountered opponent bids. To learn such a function from
experience, however, takes a long time. In fact, that time will be too
long if the other population settles more quickly at some value that
is advantageous for it. Therefore, it makes sense to do some
preprocessing of the experience before trying to figure out the
response. A natural such complexity reduction consists in taking some
suitable average of the encountered opponent values. We have already
seen above that in a simple model situation, this is an asymptotically
optimal strategy. Of course, when both populations adapt at the same
time, but perhaps according to different strategies, that analysis is
no longer strictly valid, and in certain cases,
it might be better to use some weighted average, with higher weights
for the more recently encountered values. In fact, when both
populations employ such an averaging strategy, then giving higher
weights to the more recently encountered values can lead to a faster
convergence to the optimal value.\\
Although, as already mentioned, the problem here is more difficult
than the one addressed by statistical learning theory, it is
nevertheless instructive to consider how the latter would go about
it. Here, a seller would try to model the probability distribution
utilized by the buyers. The models are taken from some model class
$\Lambda$ parameterized by a parameter $\alpha$. Let $q(k\le
k_b;\alpha)$ be the probability that for the model corresponding to
$\alpha$, the value $k$ is not larger than the buyer's bid $k_b$.
The underlying assumption upon which the seller models the buyer
here is that the latter's probability distribution is stationary,
that is, does not depend on $n$. When the seller then selects his
bid $k$ according to a probability distribution $q(k;\alpha)$, his
empirical risk after encountering the buyer's bids $k_b(1),\dots
k_b(n)$ then is
\bel{5} R_{emp}(\alpha)=K-\frac{1}{n}\sum_{i=1}^n
\sum_k k\ q(k;\alpha)\ q(k\le k_b(i);\alpha). \qe
Here, he assumes
that those bids $k_b(1),\dots k_b(n)$ represent an i.i.d. sample of
the buyer's distribution. Of course, in general, this assumption is
not valid because the buyers also adapt, but we nevertheless
proceed. The seller then chooses that parameter $\alpha(n) \in
\Lambda$ for which the empirical risk $R_{emp}(\alpha(n))$ is
minimized. By the same convexity argument as before, instead of
selecting $k$ according to some distribution $q(k;\alpha)$, he finds
that it suffices to take one single value $k(\alpha(n))$, that is,
to choose a single valued distribution. His empirical risk then
becomes
\bel{6} R_{emp}(\alpha(n))=K-\frac{1}{n}\sum_{i=1}^n
k(\alpha(n))\ q(k(\alpha(n))\le k_b(i);\alpha(n)). \qe
In
particular, he will then play that value $k(\alpha(n))$ at the next
step. In principle, as a heuristic strategy, this should also be
useful in the case where his opponents are also adapting, even
though the values $k_b(i)$ then no longer represent an i.i.d.
sample.

\section{Dynamics at the population level}

As explained, we are less interested in the competition between
individual agents within a population than in the relative performance of
populations with different types of agents. This relative performance then
is gauged by the value of the equilibrium eventually reached via the repeated
interactions of the agents from the opposite populations. We can make the following simple observations:\\
When the players of one population always play the value $k_0$, then the other
populations has no choice but to adapt to that value as well. This implies that
a population whose members converge faster than the ones from the other camp to
a value that is favorable to them will be at an advantage. According to the above
analysis, an averaging strategy dampens fluctuations and thereby improves the convergence rate.
 This speed of convergence, however, does not only depend on the own strategies,
 that is, need not reflect an absolute superiority of those strategies, but also
 reflects their reactions and adaptations to the actions of the members of the other
 population. Thus, it also depends on the state of the dynamics inside the other population.
Let us consider an extreme case: The opposite population plays rather randomly, say chooses
offers from the uniform distribution of integers between 0 and $K$. The population under
consideration, however, is subject to quick and strong selection between its agents. That
means 1) that the generation length between the evolutionary steps is very short, perhaps
even $=1$. Thus, each agent has only one encounter after which his performance is evaluated.
That also means 2) that only the very best agents, perhaps only a single best one reproduces,
and his identical or slightly mutated offspring constitutes the next generation. In that case,
chances are high (and easily computed) that in each generation there is some agent in each
generation that strikes a successful deal with a very advantageous offer, for example a buyer
offering very little may by chance encounter a seller asking even less. Thus, because of the
speed of evolution assumed, our population will quickly settle around a very favorable value.
If the other population then adapts at all, it is forced to accept that value.\\
Thus, the speed of evolution in one population and the incoherent state of the other population
together lead to an equilibrium value that is very favorable to the first population. \\
In contrast, for any averaging strategy, the speed of convergence will be slower when playing
against a more random opposite camp. \\
For more complex strategies,  a more erratic state of the opposite population may also slow down
the own convergence. When a particular response needs to be created for each previous opponent
bid (as in the complex 1-round opponent strategy described below), then an erratic opponent population
forces the players to test many different options, and a player that has already created good responses
for many bids may still acquire a low fitness when by chance exposed to bid values not yet encountered
by himself or his ancestors. Conversely, when the opposite population is rather homogeneous and constant,
the players may evolve quickly to seemingly stable states, but this may hide the fact that they do not
possess adequate responses to situations that, while possible by  the rules of the game, they and their ancestors never experienced.

\section{Different strategies and simulations}

Before presenting the simulation results, we need to introduce some notations
for the system parameters and the strategies that will be used in this paper:\\
First the parameters for the evolutionary scheme of replacing a population of
players by a new one composed of possibly mutated members of the present one with a fitness based selection:
(1) generation length (time): the number of rounds played (time steps) between
two consecutive selections (if applicable);\\
(2) selection percentage: the percentage of the players who will be chosen as parents
to generate the offspring during the evolutionary process;\\
(3) mutation rate: the rate of random mutation during the evolutionary process.\\
Next we list the main strategies investigated, classified  on the basis of
the types of information they use:\\
\begin{enumerate}
\item single-number: players use no information at all; each player chooses a fixed random
offer that will be updated through the selection
\item average-previous-opponent: the average of one's opponents'
bids in the previous, say $m$ (limited and usually much
smaller than the generation length), rounds
\item for $m=1$, that strategy is called  1-round opponent: each player utilizes the offer
of his opponent in the most recent round
\item  $n$-round opponent: each player can use the offers his opponents make in the last
$n$-rounds (and not only their average) -- here, we only consider the values $n=1$
(which is the previous strategy) and 2, as otherwise, the scheme gets computationally
too complex and performs too poorly
\item average-all-previous-opponent: the average of one's opponents' bids
in all previous rounds (thus, there is no fixed $m$ here)
\item average-friend-opponent: the average of one's friends' opponents' bids
in the most recent round (here, each player has a certain number of friends within his own population)
\item average-all-friend: the average of one's friends' bids in the most recent round
(thus, here, in contrast to the previous strategies, no information about the other population is used)
\item friendship network (average-successful-friend): the average of one's
friends' successful bids in the most recent round (here, information from the other
population is used indirectly, but selectively, because their offers decide which of the friends are successful)
\end{enumerate}
For the $n$-round opponent, for $n>1$, a scheme is needed to convert the $n$ numbers remembered
into a single response. Of course, we could simply take their average, which then reduces
this to strategy 2, but we could also employ some other scheme. One possibility is to evolve a
look-up table that lists the responses for any pair of numbers between 0 and $K$. Similarly,
we can also utilize look-up tables for the other strategies, except 1, that is, instead
of simply playing the corresponding average, each agent could have a look-up table that
specifies a specific response to each number between 0 and $K$ (formally, strategy 1 is
also a special case of this, the output of the look-up table simply being reduced to a single
number that is uniformly applied to any input). Thus, a strategy comes in two variants,
a direct one and another one with look-up table. We call these two variants 'simple' and 'complex', resp.
As noted, the 2-round opponent strategy does not possess a simple version.

First of all, we would like to have a stable setting for our model.
The stability of the model may primarily depend on the values of the
system parameters,i.e., generation length, selection percentage and
mutation rate. The effect of the generation length is somewhat
dependent on the complexity of the strategy under consideration. We
will enter into this issue further later on. Generally if the
selection process takes place, accompanied by evolving the look-up
tables, then an appropriate generation length will be any number
ranging from 100 to 1000 time steps (rounds of the game). As one
knows from evolutionary optimization methods, an appropriate choice
of the selection percentage is essential. If it is too small, then
the selection will be severe and some potentially good strategies
can get eliminated too easily. If it is too large, then the
selection will be very loose and the optimization will take much
longer. According to our numerous simulations, an effective
selection percentage will be 0.5. For the random mutation rate, a
good choice will be 1 $\sim$ 5 percent and in our simulations it is
set to be 1 percent. Our population size is always 400.

\subsection{The effects of generation length}

The generation length, that is, the number of rounds played before an
evolutionary update, expresses the relation between the time allocated
to learning in the more complex strategies and the evolutionary adaptation.
Short generation length means that individual agents have little time to
improve their performance on the basis of memory and learning, but rather
are evaluated according to their short time performance. In other words,
their experience is quickly transferred to the next generation. That generation
can then explore new responses not on the basis of systematic learning, but on
the basis of random mutations. \\
We can then simply check this issue in our simulations by letting two populations
with the same strategy space, but with different generation length, play against
each other and see which one performs better. \\
For the single-number strategy, it then turns out that the minimal
generation length, 1, is optimal. This is rather obvious because a
more quickly evolving population should have an advantage. This is
confirmed by the simulation results. For more complex strategies
such as the 1- or 2-round opponent strategy or the friendship
network strategy, the situation is not so clear, as their is a
trade-off between individual improvement based on more experience
and the speed of the evolutionary search.  The simulations results
are not yet conclusive. In some simulations, there is a small chance
for the 2-round opponent strategy to gain some slight advantage over
the 1-round opponent strategy when the generation length is as long
as 20,000 steps or even longer. Note that for shorter generation
length, we have demonstrated in \cite{JostLiFirst} that the 1-round
is superior to the 2-round strategy because the latter takes too
long to evolve.  Further simulations with very long generation times
are  required.

\subsection{Efficient Information Use}

A simple analysis uses the Shannon entropy
\cite{ShannonEntropy}
\begin{equation}\label{InfoEntropy}
S=-\sum_{k=0}^{K}  p(k)\log_2 p(k),
\end{equation}
\noindent where $p(k)$ is the probability  (frequency) that
offer $k$  has been chosen. In fact, in general, this will
depend on the time step $n$, and so, we should rather write $p(k,n)$ (which may approach a stationary $p(k)$ as $n
\rightarrow \infty$).\\
$S$ is maximal for a uniform distribution of the values of $k$ and
becomes 0 when only a single value of $k$ is played. In other words,
the evolution of $S$ expresses how quickly a population reaches a
unique response. One may argue that \rf{InfoEntropy} expresses
uncertainty and that therefore a fast decrease of this entropy
corresponds to a fast utilization of information. Figure 1 gives
some simulation results. In particular, populations with a fast
decrease of this entropy are more successful.

The simulations behind Fig.2 show that the complex
average-all-previous-opponent performs better than the complex
1-round opponent strategy. One could argue that this should be so
because the former utilizes more information in each step than the
latter. A simpler reason is that averaging reduces fluctuations and
therefore speeds up convergence. Further figures (Figs. 4 and 5)
demonstrate that simple strategies converge faster and perform
better than complex ones. However, the simple 1-round opponent
strategy performs poorly when compared with strategies utilizing
some nontrivial averaging. This is in line with our above
mathematical analysis.

\subsection{Ranking Different Strategies}

So far, we have investigated many different types of strategies. It
is of interest to check whether we can  consistently  rank their
performances. The basis for such a ranking is of course a pairwise
comparison, that is, we let one population, say the buyers, employ
one type of strategy and the sellers another one. Here, we need to
fix the other parameters, like generation length, mutation rate etc,
to have a ceteris paribus comparison, even though we realize that in
principle the ranking could change with different parameter values.
In the simulations presented below, generation length is 1000,
selection percentage is 0.5 and random mutation rate is 1 percent.

In \cite{JostLiFirst}, some of the complex strategies have been
compared. The 1-round opponent can beat the 2-round opponent
strategy because the former settles down to equilibrium more
quickly. The performance of the friendship network strategy is
comparable to the one of the 1-round opponent strategy,  neither of
them showing consistent superiority. The friendship network
strategies do not prefer any specific type of network topology
(keeping  the average degree of the networks fixed).

For a more systematic investigation, however, we should
 compare  different types of averaging strategies, in particular
the average-previous-opponent and  the
average-friend-opponent strategies. In the former, the average
is performed over the player's own opponents' offers in the previous,
say 5, rounds, whereas in the latter he averages his friends' opponents' bids in the
last round. In the average-friend-opponent strategy, one can include
the player's own experience,  that is, his own opponent's bid in the
last round. Because the opponents are randomly taken from the opposite
population, for sampling purposes this makes no difference, as long as
the average is computed from the same number, say 5, of bids from the
opposite camp. As the network topology might affect the speed of
convergence of the averaging scheme, this might have some effect here,
however. In any case, for a valid comparison between strategies, we assume that the
numbers of offers taken for both averaging procedures
are the same, say 5.

We first compare the simple average-previous-opponent to the simple
average-friend-opponent, that is, the strategies not employing
look-up tables, but taking the computed average directly as the next
own bid. The first one uses a somewhat larger sample, because it
takes the bid of one member of the opposite population at 5
different times, that is, it takes a spatio-temporal average,
whereas the second averages over 5 bids taken at the same time, that
is, it takes only a spatial average. In any case, both averaging
strategies quickly converge to their equilibrium, the population
average of the opposite population, typically 25. It also appears
that the average-previous-opponent converges slightly more slowly
than the average-friend-opponent, because the latter uses a more
recent sample from the opposite camp.

When we look at the complex strategies, that is, where look-up
tables are used to convert the computed average into a response,we
see from Fig. 6 that  in general the average-previous-opponent can
be slightly better than the average-friend-opponent. Thus, the
larger sample space, even though it uses partly outdated information
can yield an advantage.

We now turn to the average-successful-friend strategy. Here, each
player takes the average only of those
 offers of his friends and himself from the last round that have
 led to a successful deal. (If
none of those offers is successful, then a random offer  between 0
and $K$ is chosen.) (This average-successful-friend strategy is  the
friendship network strategy  examined in \cite{JostLiFirst}.) In the
beginning of the game, some randomness is introduced due to the low
success rate. Also apparently choosing only successful offers for
averaging makes the players more and more timid in making their
offers. This is the reason why the average-previous-opponent
strategy is superior to the average-successful-friend strategy, both
in the simple and in the complex setting. In Fig. 7, the buyers who
use the simple average-previous-opponent strategy can achieve the
equilibrium value 14 by adapting to the sellers who use the
average-successful-friend strategy. (A coarse estimate would in fact
expect the even lower equilibrium value 12.5) Fig. 8 presents the
competition between the two strategies with
 look-up tables. Now
the difference in the performance between the two strategies is not
that large, compared to the case without  look-up tables.
This is partly due to the fact that the players are learning. We also
notice that it takes the average-previous-opponent a little longer
than the average-successful-friend to reach the equilibrium.

In the average-all-friend strategy, a player does not distinguish
whether his friends are successful or not, but simply averages all
their bids. The simple average-all-friend strategy matches the
simple average-previous-opponent strategy, with both converging to
25. This happens because both now make use of the same trivial
random distribution. Not surprisingly, the complex
average-previous-opponent strategy performs better than the complex
average-all-friend strategy, see Fig. 9. The reason, as above, is
that the former is using a larger sample space.

We have also compared the efficiency of different spatial averaging
strategies, namely, average-friend-opponent,
average-successful-friend, and average-all-friend. The comparison
between the last two is rather straightforward with
average-all-friend prevailing over average-successful-friend. The
average-friend-opponent strategy was found to be better than the
average-successful-friend strategy, with and without look-up tables.
Fig. 10 shows an example of our simulations. It is interesting to
see that the average-friend-opponent strategy is performing nearly
equivalently to the average-all-friend the performance,  without and
with look-up tables (Fig. 11). This observation again confirms that
using less selective information can be advantageous. This is not
surprising since we have already observed that the 1-round opponent
can beat the 2-round opponent strategy.

We can now rank the different strategies. The best strategy should
be simple average-all-previous-opponent. We have found that
average-previous-opponent is nearly as good as
average-all-previous-opponent when the number of rounds for
averaging is not too small. Hence the second rank consists of
complex average-all-previous-opponent and average-previous-opponent.
The third position is taken by average-all-friend and
average-friend-opponent. At the fourth position, we would put
 1-round opponent and average-successful-friend. The lowest position  belongs to 2-round opponent.
There is yet
one strategy that needs to be placed, the single-number strategy.
This last strategy does not use any information but can still be
favored in the competition with other more complex ones that use
more information.

From the above ranking we see that whether a strategy is good or bad
is not completely decided by how much information it has used.
Rather, a good strategy should be capable of processing the
information more efficiently and thus setting an early advantage as
quickly as it can. In our simulations, we find that a weighted
averaging strategy is not doing better than the normal averaging
strategy. The weighted averaging is perhaps more reasonable by
assigning the most probable offers more chances but its adaptation
also takes longer. Being simpler and more flexible is also good for
a strategy. In this regard, the simple averaging strategy that does
not need to evolve the look-up tables can beat the complex one that
does. Not to mention that the single-number strategy can beat a lot
of more complex strategies.

\section{Conclusion}
We have investigated an iterated game played between members of opposite
populations. The individual optimization leads to a population dynamics
that determines the final equilibrium reached. Equipping the members of
the two populations with different strategic options in general will
lead to an equilibrium that is more favorable to one of the two populations.
A set of strategies that is good for a population when consistently employed
by all its members incorporates quick and efficient use of the available
information without a long learning phase, that is, rather forego a careful
optimization when it takes too long to converge. Also, a population with
bolder players does better than one with more timid ones. \\
It remains to investigate the dynamics within populations more
systematically when the individual players have different strategic options.
The ones that would cause advantageous effects at the population level might
themselves be disadvantaged inside their population and therefore get eliminated
by the evolutionary scheme which then also causes a disadvantage for the population.

\newpage

Figure Captions:

Fig. 1: The time evolution of distribution of offers made by the
sellers during the first 50 generations (1000 rounds per
generation)(top panel). Here both populations use the 1-round
opponent strategy. As shown, the peak where the most favorable offer
appears becomes higher  as the evolution continues. The final value
of the peak in this figure is 300 which shows the number of players
who have bid 25. The bottom panel shows the time evolution of the
entropy related to the distribution of sellers' offers.

Fig. 2: The complex average-all-previous-opponent, where the player
bids according to all his previous encounters' offers by taking an
average, versus the complex 1-round opponent strategy, where the
player only recalls his most recent interaction ('complex' here
means developing look-up tables). In both (a) and (b), the
population with the average-all-previous-opponent strategy is doing
better. Learning is quick and the equilibrium is rather stable. The
standard deviation is calculated from 100 samples with different
realizations.

Fig. 3: The complex average-all-previous-opponent strategy versus
the 2-round opponent strategy. In both (a) and (b), the averaging
strategy is superior. Compared to the situation in Fig. (2), the
optimization takes much longer and there also exist some
fluctuations even after the equilibrium has been reached. This
occurs mainly due to the excessive information embedded in the
2-round opponent strategy.

Fig. 4: The simple- versus the complex average-all-previous-opponent
strategy. The simple averaging strategy is superior to the complex
one. The median fee, namely the gap, is still somewhat high in both
(a) and (b). This happens mainly because the players who employ the
simple averaging strategy can converge more quickly and set an early
advantage. The players who use the complex averaging strategy have
to adapt accordingly and are forced into a disadvantageous
situation. Learning is quick because of the convergence due to the
simple averaging.

Fig. 5: Comparison between the simple average-all-previous-opponent
and the complex 1-round opponent strategy.  The simple
average-all-previous-opponent is  better than the complex 1-round
opponent.

Fig. 6: The complex average-previous-opponent versus the complex
average-friend-opponent strategy. The former can do slightly better
than the latter, mainly because the former has a slight larger
sample by doing spatio-temporal averaging and the latter is only
doing spatial averaging.

Fig. 7: The simple average-previous-opponent versus the simple
average-successful-friend strategy. The former is superior.
Distinguishing the friends by success makes the players too
cautious.

Fig. 8: The complex average-previous-opponent versus the complex
average-successful-friend strategy, with the former defeating the
latter. Here the advantage is not significant as  in Fig. 7,
primarily due to effective learning.

Fig. 9: The complex average-previous-opponent versus the complex
average-all-friend strategy, with the former being slightly better.

Fig. 10: The comparison between the complex average-friend-opponent
and the complex average-successful-friend strategy. The former is
better, which indicates again that selecting only successful friends
makes the players too timid.

Fig. 11: The complex average-friend-opponent is almost equivalent to
the complex average-all-friend strategy. Compared to what has been
shown in Fig. 11, using less selective information is not inevitably
disadvantageous.

%Fig. 12: The performance of individual erratic agents versus the
%degree of randomness. Here the degree of randomness is represented
%by the standard variance $s$ of a Gaussian distribution. The
%performance is evaluated by the average score of the fellow players
%holding the same $s$ within a certain population.

%Fig. 13: The performance comparison between the buyer population
%whose members use 1-round opponent strategy and the seller
%population whose most members also use 1-round opponent strategy but
%5 percent of them behave randomly. The curve shows the dominance of
%the sellers.

\end{document}